\documentclass[11pt]{article}
\usepackage{epsf,amssymb,cite}
\newcommand{\xtra}[1]{{.}}
\renewcommand{\xtra}[1]{{, \tt hep-th/#1.}}

\newcommand{\cH}{{\cal H}}
\newcommand{\One}{\hbox{{\rm 1{\hbox to 1.5pt{\hss\rm1}}}}}

\newcommand{\nn}{\nonumber}

\newcommand{\ri}{\right}
\newcommand{\lf}{\left}

\newcommand{\Ga}{\Gamma}

\newcommand{\ep}{\varepsilon}
\newcommand{\eq}{\begin{equation}}
\newcommand{\en}{\end{equation}}
\newcommand{\bea}{\begin{eqnarray}}
\newcommand{\eea}{\end{eqnarray}}

\newcommand{\ba}{\begin{array}}
\newcommand{\ea}{\end{array}}

\newcommand{\CC}{{\hbox{\rm C\kern-0.5em{$\sf I$}}}}
\newcommand{\II}{\hbox{{\rm l{\hbox to 1.5pt{\hss\rm l}}}}}
\newcommand{\RR}{{\hbox{$\rm\textstyle I\kern-0.2em R$}}}
\newcommand{\ZZ}{{\hbox{$\sf\textstyle Z\kern-0.4em Z$}}}

\newcommand{\resection}[1]{\setcounter{equation}{0}\section{#1}}

\newcommand{\NP}[1]{Nucl.\ Phys.\ {\bf #1}}
\newcommand{\PL}[1]{Phys.\ Lett.\ {\bf #1}}

\newcommand{\PRL}[1]{Phys.\ Rev.\ Lett.\ {\bf #1}}

\newcommand{\IJMP}[1]{Int.\ J.\ Mod.\ Phys.\ {\bf #1}}

\newcommand{\TMP}[1]{Teor.\ Math.\ Phys.\ {\bf #1}}
\newcommand{\zm}{Zamolodchikov}
\newcommand{\ABZ}{A.B.~Zamolodchikov}
\newcommand{\AlBZ}{Al.B.~Zamolodchikov}

\newcommand{\JP}[1]{J.\ Phys.\ {\bf #1}}

\newcommand{\usbl}[1]{\left(#1\right)}
\newcommand{\inti}{\int^{\infty}_{-\infty}}
\newcommand{\iintd}{\int^{\infty}_{-\infty}\! d\theta}

\newcommand{\ket}[1]{|#1\rangle}
\newcommand{\bra}[1]{\langle #1|}

\newcommand{\MR}{M\! R}

\newcommand{\fract}[2]{{\textstyle\frac{#1}{#2}}}

\newcommand{\prtial}{\frac{\partial}{\partial\theta}}

\newcommand{\opnup}[1]{\renewcommand{\\}{\\[50 pt]}}
\renewcommand{\bar}{\overline}

\newcommand\re{\hbox{Re}\,}
\newcommand\im{\hbox{Im}\,}

\newcommand\LYM{{\cal M}(2/5)}
\newcommand\Ebndry{E_{\rm bndry}}
\newcommand\Eblk{{\cal E}_{\rm bulk}}
\renewcommand\hat{\widehat}
\begin{document}
%
\begin{titlepage}
\vskip 0.5cm
\begin{flushright}
DTP-97-67 \\
KCL-MTH-97-71 \\
SPhT-97/163 \\
{\tt hep-th/9712197}\\
December 1997 \\
(revised)
\end{flushright}
\vskip 1.2cm
\begin{center}
{\Large {\bf TBA and TCSA with boundaries and}} \\[5pt]
{\Large {\bf excited states } }
\end{center}
\vskip 0.8cm
\centerline{Patrick Dorey%
\footnote{e-mail: {\tt P.E.Dorey@durham.ac.uk, 
A.J.Pocklington@durham.ac.uk}}, Andrew Pocklington$^1$, 
Roberto Tateo\footnote{e-mail: {\tt Tateo@wasa.saclay.cea.fr}}
and G\'erard Watts\footnote{e-mail: {\tt gmtw@mth.kcl.ac.uk}}
}
\vskip 0.6cm
\centerline{${}^1$\sl Department of Mathematical Sciences,}
\centerline{\sl  University of Durham, Durham DH1 3LE, 
England\,}
\vskip 0.2cm
\centerline{${}^2$\sl Service de Physique Th\'eorique, CEA-Saclay,}
\centerline{\sl F-91191 Gif-sur-Yvette Cedex, France\,}
\vskip 0.2cm
\centerline{${}^3$\sl Mathematics Department, }
\centerline{\sl King's College London, Strand, London WC2R 2LS, U.K.}
\vskip 0.9cm
\begin{abstract}
\vskip0.15cm
\noindent
We study the spectrum of the scaling Lee-Yang model on a finite interval
{}from two points of view: via a generalisation of the truncated
conformal space approach to systems with boundaries, and via the
boundary thermodynamic Bethe ansatz. This allows reflection factors to
be matched with specific boundary conditions, and leads us to propose
a new (and non-minimal) family of reflection factors to describe the
one relevant boundary perturbation in the model. The equations
proposed previously for the ground state on an interval must be
revised in certain regimes, and we find the necessary modifications by
analytic continuation. We also propose new equations to describe
excited states, and check all equations against boundary truncated
conformal space data. Access to the finite-size spectrum enables us to
observe boundary flows when the bulk remains massless, and the
formation of boundary bound states when  the bulk is massive.\\
\end{abstract}
\end{titlepage}
\setcounter{footnote}{0}
\def\thefootnote{\fnsymbol{footnote}}

\resection{Introduction}
Integrable quantum field theories in domains with
boundaries
have attracted some attention of late, principally in the
wake of a paper by Ghoshal and \zm~\cite{GZa}. Such
theories can be specified in terms of ultraviolet data consisting of
the original boundary conformal field theory together with a
specification of the particular perturbations chosen, both in the bulk
and/or at the boundary, or else
in terms of infrared data, which might
consist of a bulk S-matrix together with a set of reflection factors
encoding the scattering of each particle in the model off a single
boundary. The reflection factors are constrained by various
consistency conditions -- unitarity,
crossing-unitarity~\cite{GZa}, and the boundary Yang-Baxter~\cite{Ca}
and boundary bootstrap~\cite{FKa,GZa}
equations -- and it is natural to explore the space
of solutions to these conditions, and then attempt to match them with
particular perturbed boundary conformal field theories. While the
first aspect has been studied by many authors,
the second is comparatively underdeveloped, and provides much of the
motivation for the work to be described in this paper. The strategy
(already employed with much success in boundaryless situations) will
be to study finite-volume spectra by two different routes, one based on
the ultraviolet data and one on the infrared data, and then to compare the
results.  The ultraviolet route will proceed via a generalisation of
the truncated conformal space approach (TCSA)~\cite{YZa} to boundary
situations. This seems to us to be of independent interest, and should
be useful in non-integrable situations too. For brevity, we shall
refer to this method as the BTCSA, for boundary truncated conformal
space approach. On the infrared side, the
main weapon will be the `BTBA', a modification of the thermodynamic Bethe
ansatz (TBA) equations~\cite{Zb} adapted to boundary situations,
introduced 
in refs.~\cite{Zta,LMSSa}. We have made a study of the
analytic structure of these modified equations, uncovering a number of
surprises along the way. We have also found new equations, some of
them encoding
the ground state energy in regimes where the
previously-proposed equations break down, and others
the energies of excited states. These new equations are similar
in form to those found in~\cite{BLZa,DTa} for excited states in the
more traditional (boundaryless) TBA.  The agreement between
BTCSA and BTBA results offers support for both, and in addition allows
us to obtain concrete predictions about the relationship between
various parameters appearing at short and long distances.

Thus far, most of our work has been confined to what appears to
be the simplest 
nontrivial example, namely the boundary scaling
Lee-Yang model, but the methods advocated certainly have wider
applicability. In this paper the emphasis will be on the structure of
the BTBA equations and the matching of their solutions with the
results that we have obtained by means of the BTCSA. A more detailed
description of the BTCSA method itself will appear in a companion
paper, currently in preparation~\cite{Us2}.

\resection{The model}
%
%
The ultraviolet limit of the scaling Lee-Yang model (SLYM) is the 
$\LYM$ minimal
model~\cite{Cardy85}, a non-unitary conformal field theory
with central charge $c = -22/5$.
The field content and the operator description depends on
the geometry considered, but far from any boundary the local fields
are those of the theory on a plane, and are left and right
Virasoro descendents of
the primary fields
$\One$, of scaling dimension $0$, and $\varphi$, of scaling
dimension $x_\varphi =\Delta_{\varphi}{+}\overline\Delta_{\varphi}=
 -2/5$. Both of these are scalars, and the one
non-trivial fusion rule is
$\varphi \times \varphi = \One + \varphi$.
The conventional normalisation of $\varphi$ is
\eq
  \varphi(z) \; \varphi(w) 
= |z-w|^{4/5} \;+\; C_{\varphi\varphi}^\varphi\,\varphi(w)\,|z-w|^{2/5} + \ldots\;,
\label{eq:norm1}
\en
but this results in $C$ being purely imaginary, and many other
structure constants in the boundary theory being imaginary as
well. For this reason, we have chosen the non-standard normalisation
\eq
  \varphi(z) \; \varphi(w) 
= - |z-w|^{4/5} \;+\; C_{\varphi\varphi}^\varphi\,
    \varphi(w)\,|z-w|^{2/5} + \ldots\;,
\label{eq:norm2}
\en
where we take 
$C_{\varphi\varphi\varphi} {=} - C_{\varphi\varphi}^\varphi$ 
to be a positive real number.

If the physical geometry has a boundary, then a
conformally invariant boundary condition (CBC) must be assigned
to each component of that
boundary, and the possible local fields on any part of
the boundary depend on the particular conformal boundary condition
found there. Cardy has classified the possible boundary
conditions~\cite{Cardy89} and the field
content on each of these can be found
by solving the consistency conditions given
by Cardy and Lewellen~\cite{CL91}. The boundary fields fall into
irreducible representations of a single Virasoro algebra, so all that
is needed to
specify the local field content of a particular conformally invariant
boundary condition is the set of weights of the primary
boundary fields.
 
It turns out that there are two conformal boundary conditions
for $\LYM$, which by an abuse of notation we shall
label by $\One$ and $\Phi$. The $\One$ boundary has only one primary
boundary field, which is the identity $\One$, while the $\Phi$ boundary
has two, the identity and a field $\phi$ of
scaling dimension $x_\phi = -1/5$.
As a result, there are no relevant boundary perturbations of the
$\One$
boundary, and a single relevant perturbation of the $\Phi$ boundary,
by the field $\phi$.
As with equation (\ref{eq:norm2}), we choose the normalisation of
$\phi$ to be 
\eq
  \phi(x) \; \phi(y) 
= - |x-y|^{2/5} \;+\; C_{\phi\phi}^\phi\,
    \phi(y)\,|x-y|^{1/5} + \ldots\;,
\label{eq:norm3}
\en
with 
$C_{\phi\phi\phi}^{\vphantom{\phi}}= - C_{\phi\phi}^\phi$ a positive
real number.
Together equations (\ref{eq:norm2}) and (\ref{eq:norm3})
determine the bulk-boundary constant ${}^{\Phi}\!B_{\varphi}^\phi$
appearing in the expansion of $\varphi(x+iy)$ on the upper half plane
with boundary condition $\Phi$ at $y=0$:
\eq
  \varphi(x+iy)
= {}^{\Phi}\!B_{\varphi}^\phi\,\phi(x)\,(2y)^{1/5} + \ldots\;.
\en
The only remaining independent structure constant appears in the
operator product expansion
of two boundary-changing operators, but is not needed to reproduce
the results in this paper. 

This sketch will suffice for 
a description of the various boundary scaling Lee-Yang models that
we shall be considering. Each can be regarded as a perturbation of one
of the boundary conformal field theories just discussed.
 

First, suppose that there is no boundary at all. There is only the
bulk to perturb, and if the perturbation is to be relevant then there
is only one bulk field, namely $\varphi$, to perturb by. 
The perturbed action
\eq
{\cal A}_{\rm SLYM}={\cal A}_{\LYM}+
\lambda\!\inti\!dy\inti\!dx\,\varphi(x,y) 
\label{arnold}
\en
is integrable, and, for $\lambda>0$, results in a massive scattering
theory with a single particle type of mass $M$, and two-particle
S-matrix~\cite{CMa}
\eq
S(\theta)=-\usbl{1}\usbl{2}~~,\quad
\usbl{x}={\sinh\bigl({\theta\over 2}+{i\pi x\over 6}\bigr)\over
        \sinh\bigl({\theta\over 2}-{i\pi x\over 6}\bigr)}~.
\label{asm}
\en
The exact relationship between $M$ and $\lambda$ was found in~\cite{Zg}.
In the conventions implied by (\ref{eq:norm2}) and (\ref{arnold}), it is
\eq
M(\lambda)=
{2^{19/12}  \sqrt{\pi} \over 5^{5/16}}  {\lf( \Ga(3/5) \Ga(4/5)
\ri)^{5/12}\!\!\!\!\!\!\!\!\!
\over
\Ga(2/3) \Ga(5/6) }\;\;\;\;\;\lambda^{5/12}
= (2.642944\dots)\lambda^{5/12}\;. 
\label{mlrel}
\en
Now add a single boundary along the imaginary axis $x=0$. Then, as
explained in, for example,~\cite{GZa}, 
the S-matrix should be supplemented with a
reflection factor encoding how the particle bounces off the boundary.
There are four `minimal' possibilities, each of which satisfies
all of the consistency conditions entailed by the S-matrix
while minimising the number of poles and zeroes in
the strip $0\le\im\theta\le\pi\,$\footnote{It might be more
natural to minimise the number of poles and zeroes in the narrower
strip $0\le\im\theta\le\pi/2\,$. In any event,
we won't be imposing
either version of minimality, but rather checking solutions directly
against BTCSA data.}:
\bea
R_{(1)}=  
\usbl{\fract{1}{2}}\usbl{\fract{3}{2}}\usbl{\fract{4}{2}}^{-1}
\phantom{-}
&,&~\quad R_{(2)}=
\usbl{\fract{3}{2}}^{-1}\!\usbl{\fract{4}{2}}^{-1}\!
\usbl{\fract{5}{2}}^{-1}\nn\\
R_{(3)}=  
-\usbl{\fract{1}{2}}\usbl{\fract{2}{2}}\usbl{\fract{3}{2}}
{}~~\,&,&~\quad R_{(4)}= 
-\usbl{\fract{2}{2}}\usbl{\fract{3}{2}}^{-1}\!\usbl{\fract{5}{2}}^{-1}
\label{lew}
\eea
The first two were given in~\cite{GZa}, while the second
two are related to these by multiplication by the bulk S-matrix.
(As observed in~\cite{Sa}, this procedure
automatically generates further solutions
to the consistency conditions.) 
In fact, an easing of the minimality requirement will be required
in order to match all of the boundary conditions that will be encountered. 
For the time being we just note
that the reflection factor
\eq
R_b(\theta)=
\usbl{\fract{1}{2}}
\usbl{\fract{3}{2}}
\usbl{\fract{4}{2}}^{-1}\!
\usbl{\fract{1-b}{2}}^{-1}\!
\usbl{\fract{1+b}{2}}
\usbl{\fract{5-b}{2}}
\usbl{\fract{5+b}{2}}^{-1}
\label{eq:rfb}
\en
is consistent with the bulk S-matrix for any value of the parameter
$b$, and reduces to 
$R_{(1)}$, $R_{(2)}$ and $R_{(3)}$ for $b=0$, $-2$ and $1$
respectively. 

At this stage there is no way of telling which, if any, of
these solutions is actually realised as the reflection factor of a
concrete perturbed boundary conformal field theory. 
Such a theory, in the geometry currently under consideration,
will have an action of the form
\eq
{\cal A}_{\rm BSLYM}={\cal A}_{\LYM+{\rm CBC}}+ 
\lambda\!\inti\!dy\int_{-\infty}^0\!dx\,\varphi(x,y)+
h\!\inti\!dy\,\phi_B(y)~,
\label{arnie}
\en
where ${\cal A}_{\LYM+{\rm CBC}}$ is the action for the conformal
field theory on the semi-infinite plane, with a definite
conformal boundary condition at $x=0$, and $\phi_B(y)$ is one of
the boundary fields allowed by that same boundary condition.
We shall denote the boundary $\Phi$ with a term in the action
$ h\!\oint ds\,\phi(s)$
by $\Phi(h)$. It is important to appreciate that since the bulk-boundary
coupling ${}^{\Phi}\!B_{\varphi}^{\phi}$ is non-zero, the sign of $h$
is important, just as the fact that the bulk three-point
coupling is non-zero means that the sign of
$\lambda$ in  (\ref{arnold}) and (\ref{arnie}) is important.
 
As a final step, we add a second boundary to confine the system to a
strip of finite width. The next section outlines how this situation
can be analysed numerically, using a modification of the TCSA technique.

\resection{Boundary TCSA}
%
%
The TCSA method of~\cite{YZa} assumes periodic boundary conditions,
and has to be adapted
to our context of a strip of width $R$.
A conformal mapping from the strip to the upper half plane sends
the quantisation surface to a semi-circle of radius $1$, with the 
left and right boundary-perturbing operators sitting on the real axis 
at $-1$ and $1$ respectively.
The Hamiltonian is then given in terms of the boundary conformal field
theory as
\bea
  H_{\alpha\beta}(M,R)
&=& \frac{\pi}{R}
  \, \left[\;
   L_0  - \frac{c}{24}
 \;+\;
  \lambda
  \left( \frac{R}{\pi}\right)^{2 - x_\varphi}
  \!\!\!\int_{0}^\pi\!\!d\theta\, \varphi( \exp(i \theta))
  \right.
\nn \\[3pt]
&& \;\;\;\;\;\;\;\;\;\;+\;
\left.
  h_L
  \left(\frac{R}{\pi}\right)^{1-x_{\rm L}}\!\phi_{\rm
L}(-1)
 \;+\;
  h_R
  \left(\frac{R}{\pi}\right)^{1-x_{\rm R}}\!\phi_{\rm
R}(1)
 \;\right]
\,,~~~~~
\label{eq:tcsa1}
\eea
where $\varphi$ is the bulk primary field and $\phi_{\rm L}$
and $\phi_{\rm R}$ are the boundary perturbations (if any)
applied to the left and right boundaries of the strip. The scaling
dimensions of these operators are $x_{\varphi}$, $x_{\rm L}$ and $x_{\rm
R}$ respectively.
The subscripts $\alpha$ and $\beta$ of $H_{\alpha\beta}(M,R)$
are included as reminders of the information residing in
the left and right boundary conditions and
perturbations, while $M$ is related to the bulk coupling $\lambda$
according to the relation (\ref{mlrel}).
Since we are dealing here with relevant perturbations of
the scaling Lee-Yang model,
$x_{\varphi}$ is equal to $-2/5$, and each of
$\phi_{\rm L}$ and $\phi_{\rm R}$ is either absent, 
or is the boundary field $\phi$ with scaling dimension
$x_{\phi}=-1/5$. 
This latter option only arises when perturbing $\Phi$ boundaries -- as
mentioned in the last section, the $\One$ boundary does not support
any relevant boundary perturbations.
 
The essence of the BTCSA is to diagonalise the Hamiltonian
(\ref{eq:tcsa1}) in a finite-dimensional subspace of
the full Hilbert space, usually obtained by discarding those states of
conformal weight larger than some cutoff.
The matrix elements of the fields 
$\phi_{\rm L}(-1)$,
$\phi_{\rm R}(1)$
and
$\varphi(\exp(i\theta))$ can be found exactly, but in general the
integral over $\theta$ in (\ref{eq:tcsa1}) has been performed
numerically. 

If the Hilbert space of the model on a strip with conformal boundary
conditions $\alpha$ and $\beta$ is
$\cH_{(\alpha,\beta)}\,$, and the 
irreducible Virasoro representation of weight $h$ is $V_h$, then 
using~\cite{Cardy89} we found
\eq
  \cH_{(\One,\One)} = V_0
\;,\;\;\;\;
  \cH_{(\One,\Phi)} = V_{-1/5}
\;,\;\;\;\;
  \cH_{(\Phi,\Phi)} = V_0 \oplus V_{-1/5}
\;.
\label{rowlf}
\en
The full set of structure constants and correlation functions
necessary to evaluate (\ref{eq:tcsa1}) on these spaces can then
be found using \cite{CL91}, and will be given in \cite{Us2}.
 
We used the BTCSA to investigate the model with a pure
boundary perturbation ($\lambda {=} 0$), and also with
combined bulk / boundary perturbations.
In the first case, 
perturbing by a relevant boundary field 
while leaving the bulk massless
gives a renormalisation group
flow which, while leaving the bulk properties of the conformal field theory
unchanged, moves from one
conformal boundary condition to another.
The simplest context where we can study this phenomenon within the
BTCSA is the bulk-conformal Lee-Yang model on a strip with
$(\One,\Phi)$ boundary conditions. Perturbing the $\Phi$ boundary by
the $\phi$ field should provoke a flow to another conformal boundary
condition with fewer relevant boundary operators. The only candidate
is $\One$, and that is indeed what we 
found for {\em positive} values of $h\equiv h_R$. 
Since the bulk is massless, the only length scale in the
problem is provided by the boundary field $h$, and the 
crossover occurs as a function of
the dimensionless quantity $r=h^{5/6}R$.
The best signal is provided by a plot of scaling
functions, 
which we define in terms of the energy levels $E_n(h,R)$
by
\eq
E_n(h,R)=
\frac{\pi}{R}F_n(h^{5/6}R)~.
\label{polenta}
\en
In figure 1a we display the gaps $(F_n(r) - F_0(r))$ for the
$(\One,\Phi(h))$ boundary conditions with $\log(r^{6/5}){=}\log(h
R^{6/5})$ varying from $-4.9$ to $2.7$, using the BTCSA truncated to 98
states. We have used a logarithmic scale for to highlight the
crossover region, in which there is a smooth flow between the
$\cH_{(\One,\Phi)}$ and $\cH_{(\One,\One)}$  spectra. As should be
clear from the plot, the levels move from the degeneracy pattern given
by $\chi_{-1/5}\,$, the character of the $V_{-1/5}$ representation, to
that given by $\chi_0\,$: 
\eq
\begin{array}{ll}
\!q^{1/60}\,\chi_{{-}1/5}(q) = \\[2pt]
{}~~
 1 + q + {q^2} + {q^3} + 2\,{q^4} + 2\,{q^5} + 3\,{q^6} + 3\,{q^7} + 
  4\,q^8 + 5\,q^9 + 6\,q^{10} + 7\,q^{11} 
 + O(q^{12})\;,~~~~
\\[7pt]
\!q^{-11/60}\;\chi_{0}(q)\,= \\[2pt]
{}~~
 1 ~~~+~~ {q^2} + {q^3} ~+~{q^4} ~+~ {q^5} + 2\,{q^6} + 2\,{q^7} +
         3\,q^8 + 3\,q^9 + 4\,q^{10} + 4\,q^{11}
 + O(q^{12})\;.
\end{array}
\en
With $h$ {\em negative}\ the energies become complex
at large (real) $R$, just as happens in the bulk if $\lambda$ is
negated~\cite{YZa}. However, in this case the spectrum remains
identifiable. 
For large negative $h$, it splits into
three components. There are two complex conjugate sets, the real parts of
which are counted by the character $\chi_0(q)$; and
a third set which is real and which is also counted by the character
$\chi_{0}(q)$, but which appears to have a 
different asymptotic behaviour in $h$ and to decouple
in the $h\to\infty$ limit.
To illustrate this point, in figure 1b 
we plot the first 53 scaling functions against $r^{6/5}=hR^{6/5}$,
for $-12.2<hR^{5/6}{<}7.4$. 
The solid lines indicate real eigenvalues, and the dashed lines the real
parts of complex eigenvalues.
At the moment we are unsure quite what the interpretation of these
complex eigenvalues is, or why the different components
should be organised into representations of the Virasoro algebra.
However similar  spectra have been found before, for example see
ref.~\cite{BWeh1}. 

{\begin{figure}[ht]
\[\begin{array}{ll}
\epsfxsize=.47\linewidth\epsfbox{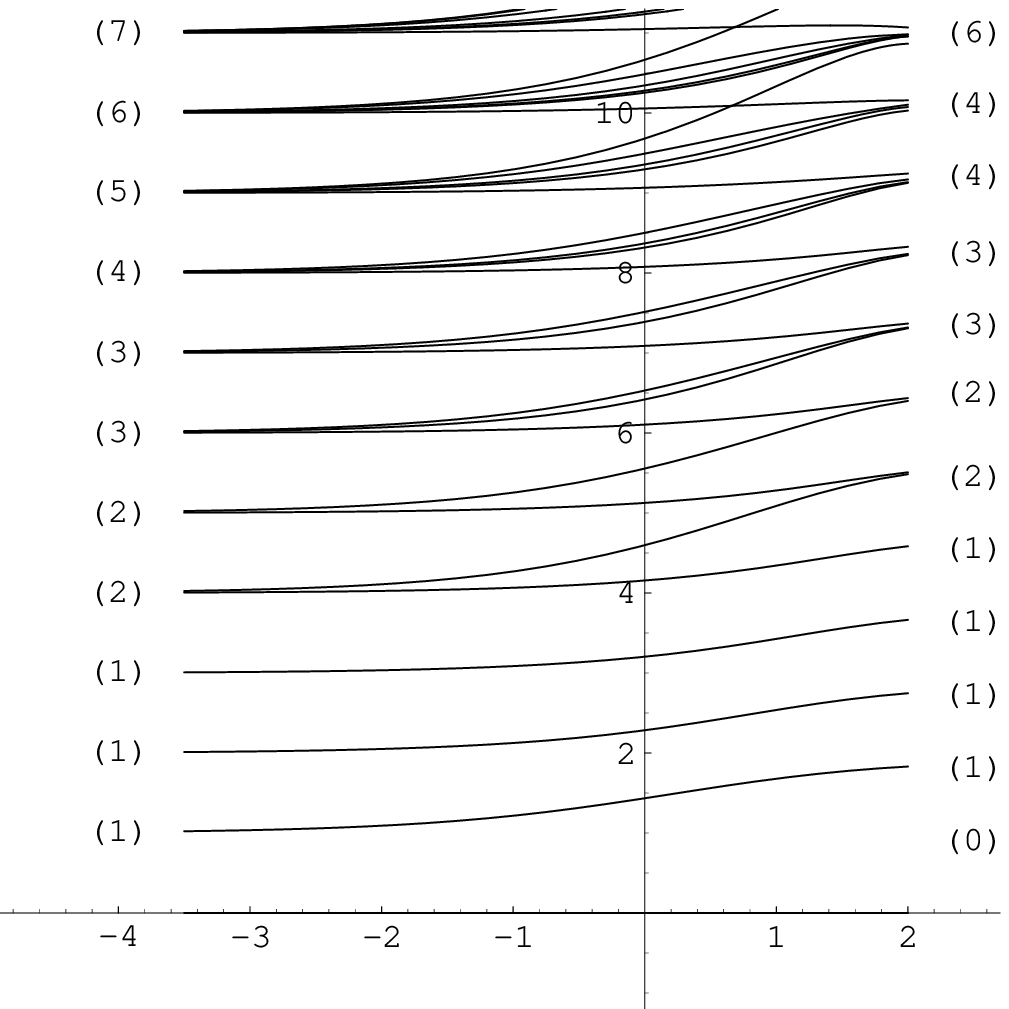}
{}&
\epsfxsize=.47\linewidth\epsfbox{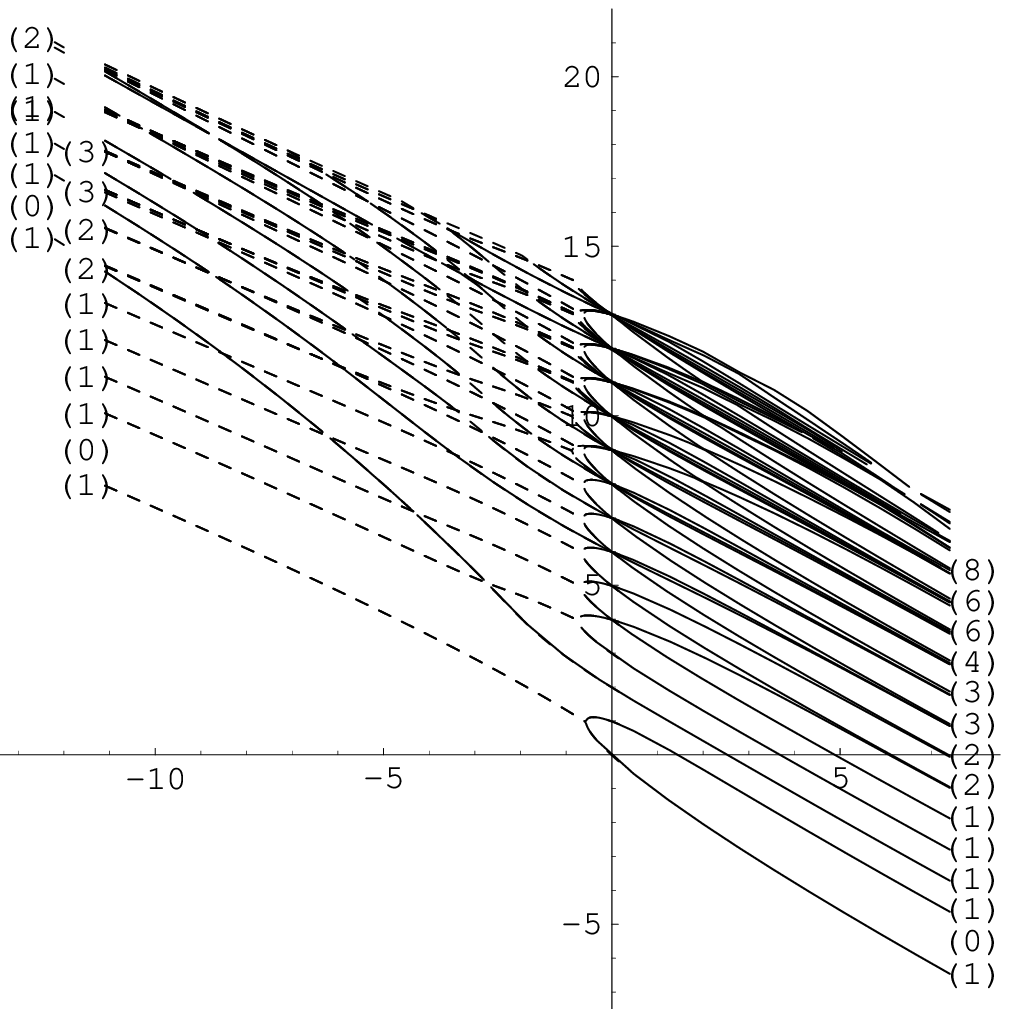}
\\
\parbox[t]{.47\linewidth}{\small 1a)
The excited state scaling function gaps as a function of $\log(
r^{6/5})$ for the $M{=}0$ LY model on a strip with boundary conditions
$(\One,\Phi(h))$, with $h=(r/R)^{6/5}$. The multiplicities are in
parentheses.}  
{}~&~
\parbox[t]{.47\linewidth}{\small 1b)
The first 53 scaling functions $F_n(r)$ plotted against $r^{6/5}$ for
the $M{=}0$ LY model on a strip with boundary conditions
$(\One,\Phi(h))$, with $h=(r/R)^{6/5}$. The multiplicities are in
parentheses.} 
\end{array}\]
\end{figure}}

We have also used the BTCSA to investigate massless
boundary flows for other unitary and non-unitary minimal models, with
both integrable and non-integrable boundary perturbations.
In unitary cases the spectrum remains real, but in other respects
the picture just outlined seems to be rather general.
More work will be needed before
the full picture is clear, but one particular point is worth
stressing:  if the bulk is {\it massless}, then a 
boundary perturbation cannot affect this, and hence {\it any}
boundary perturbation, integrable or non-integrable,
must flow
to a conformal, and indeed integrable, boundary condition under the
renormalisation group. This contrasts with the generic behaviour of a
bulk perturbation, where fine-tuning 
is required if the infrared limit is to be anything other than
massive.  It also helps to explain why we were able to observe massless
boundary flows in the BTCSA, despite the errors caused by
the truncation to a finite number of levels. In analogous bulk
situations, the TCSA makes a rather bad job of modelling bulk-massless 
flows, even in situations where the oppositely-perturbed, bulk-massive, 
flows are captured fairly well -- see for example section 5.2 
of~\cite{LMCa}.

\medskip

As a second application of the BTCSA, 
we have investigated combined bulk / boundary
perturbations in order to unravel the connection between the
reflection factors (\ref{lew}), (\ref{eq:rfb}) and particular
perturbed boundary conformal field theories.
To this end it is only necessary to consider
the $(\One,\One)$ and $(\One,\Phi(h))$ boundary conditions, but
we defer the presentation
of these results until later in the paper, after the discussion of
the BTBA method.

\medskip

Finally, we have used the BTCSA to investigate the complete partition
function of the scaling Lee-Yang model
 on a cylinder of length $R$ and circumference
$L$, with bulk mass $M$ and general
boundary conditions $(\alpha,\beta)$ at the two ends of the
cylinder. In terms of the spectrum of the strip Hamiltonian
(\ref{eq:tcsa1}), this partition function is 
\bea
  Z_{\alpha\beta}(M,R,L)
&=&{\rm Tr}_{\cH_{(\alpha,\beta)}} e^{-LH_{\alpha\beta}(M,R)}\nn\\[5pt]
&=&\!\!\!\!
 \sum_{E_n\in\,{\rm spec}(H_{\alpha\beta}(M,R))}\!\!\!\!\!\!
 \exp( - L E_n )
\;.
\label{bucket}
\eea
This partition function can also be evaluated by propagating states along
the length of the cylinder, giving the large-$R$ asymptotic
\eq
 Z_{\alpha\beta}(M,R,L) \sim_{R \to \infty}\,
 A_{\alpha\beta}(M,L)\, \exp( - R \,{E^{\rm p}_0(M,L)} )
\label{lrasympt}
\en
where $E^{\rm p}_0(M,L)$ is the ground state energy of the model
on a circle of circumference $L$. 
The linear part of
$\log A_{\alpha\beta}$ can be extracted by setting 
\eq
  \log(A_{\alpha\beta}(M,L))
=\log(\,g_{\alpha}(M,L)\,g_{\beta}(M,L)\,)
  + B_{\alpha\beta}\, L
\;,
\label{eq:aggb2}
\en
For the two boundary conditions we have, the two functions
$g_{\One}(M,L)$ and $g_{\Phi(h)}(M,L)$ can be expressed in terms of
dimensionless combinations as
\eq
  g_{\One}(M,L) = f_1(\, M\!L\,)
\;,\;\;\;\;
  g_{\Phi(h)}(M,L) = f_2(\, M\!L \,,\, h L^{6/5} \,)
\;.
\en
The numbers $f_1(0)$ and $f_2(0,0)$
are the ``universal ground state degeneracies" discussed in~\cite{AL}
of the corresponding UV conformal boundary conditions, i.e. $\One$ and
$\Phi(0)$ respectively, with values as given in table 
\ref{tab:lgg}.
The opposite, $L\to\infty$, limit 
must be taken
with more care, as results are different for the two cases of massless
and massive bulk.

At $M{=}0$ (critical bulk), for $h>0$ the function $f_2(0, h L^{6/5})$
governs the crossover between conformal boundary conditions 
$\Phi(0)$ and $\One$, so that we expect \cite{AL}
\eq
  \lim_{L \to \infty\atop h>0} g_{\Phi(h)}(0,L) = 
  \lim_{\kappa \to +\infty} f_2(0,\kappa ) = f_1(0)
\;.
\label{eq:lim1}
\en
However, for a massive bulk, we expect that the limit $L \to \infty$ 
will lead to a purely massive boundary theory, for which the ground
state degeneracy is 1, so that in all cases for the 
Lee-Yang model,
\eq
  \lim_{L \to \infty\atop M>0} g_{\alpha}(M,L) = 1
\;.
\label{eq:lim2}
\en

The quantities that we have just been discussing arise in a limit
where the formula (\ref{bucket}) might be expected to break down, at
least when truncated to a finite number of levels. Nevertheless, it
turned out to be possible to extract both $E^{\rm p}_0(M,L)/M$ and the
product $g_{\alpha}(M,L)\,g_{\beta}(M,L)$ from the numerical
approximation to the partition function as calculated from the
spectrum of the BTCSA  Hamiltonian. 

As a first example, we calculated $\log|E^{\rm p}_0(M,L)/M|$
from the BTCSA approximation to the partition function, and
compared it with the result obtained directly by using the TCSA for a
circle. The result should be independent of the boundary conditions,
and that is indeed what we found. The results are given in figure 2a, 
in which we compare the direct TCSA (or TBA) evaluation of
$\log|E^{\rm p}_0(M,L)/M|$ for the model
on a circle (solid line) with the result obtained from 
the BTCSA evaluation of
$Z_{\One\Phi(0)}$ with a truncation to 29 states
(points), and of $Z_{\One\One}$ with a
truncation to 26 states (open circles). 
As can be seen, the values for 
$E^{\rm p}_0(M,L)/M$ obtained via the BTCSA are indeed independent of
the boundary conditions on the strip, and agree with the expected
answers obtained by looking directly in the other channel. 

{\begin{figure}[htb]
\[\begin{array}{c}
{\epsfxsize=.57\linewidth\epsfbox[0 50 250 210]{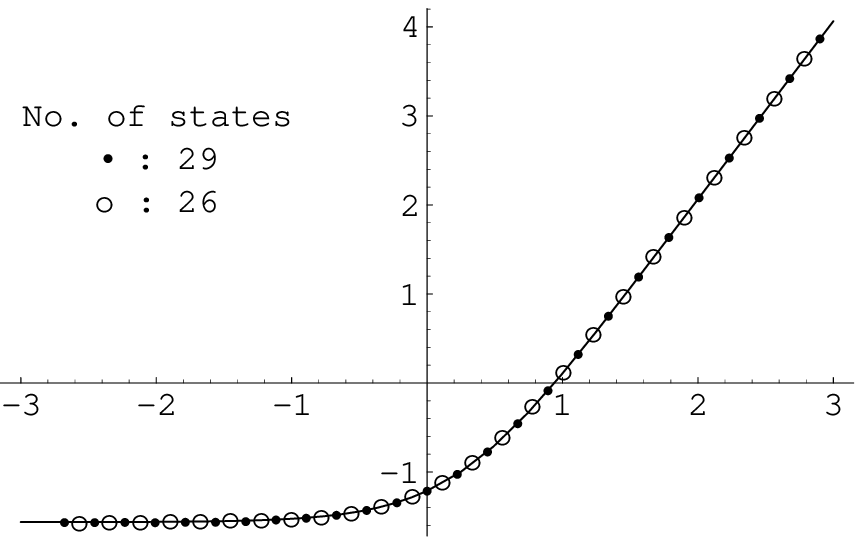}}\\
\parbox[t]{.57\linewidth}{\small 2a)
$\log |E^{\rm p}_0(M,L)/M|$ plotted against $\log(ML)$ for the SLYM on
a circle: TCSA for the system on a circle (line) compared with
results from the BTCSA with boundary conditions
$(\One,\Phi(0))$ (points) and $(\One,\One)$ (circles).}
\end{array}\]
\end{figure}}

For our second example we estimated the ground state degeneracy functions
$\log(\, g_{\alpha}(M,L)\, g_{\beta}(M,L)\, )$
for $(\alpha,\beta)$ taking the two values $(\One,\One)$ and
$(\One,\Phi(0))$. 
The BTCSA calculations are rather more sensitive to truncation error
than in the previous example, and we took results from various
truncations up to $98$ states, and then extrapolated in the truncation
level. The numerical results at $(M\!L)=0.02$ are compared with the exact
results at $L=0$ in table \ref{tab:lgg}.

{
\begin{table}[htb]
{\renewcommand{\arraystretch}{1.6}
\[
\begin{array}{c|c|c}
\hbox{The b.c.s} & \hbox{The exact value of} & \hbox{The BTCSA estimate
for} 
\\[-3mm]
(\alpha,\beta) & 
\log(\, g_{\alpha}(M,0)\, g_{\beta}(M,0)\, ) &
\log(\, g_{\alpha}(M,\frac{0.02}M)\, g_{\beta}(M,\frac{0.02}M)\, ) 
\\[1mm] \hline

(\One,\One) & 
\fract{1}{2}\log((5{-}\sqrt5)/10)=-0.64296... &
-0.643 \pm 0.001 \\ \hline

(\One,\Phi(0)) &
\fract{1}{2}\log((5{+}\sqrt5)/10)=-0.161754... &
-0.1616 \pm 0.0005 

\end{array}
\]}%
\vspace{-2truemm}
\caption{%
the BTCSA estimates and exact values of 
$\log(\, g_{\alpha}(M,0)\, g_{\beta}(M,0)\, )$}
\label{tab:lgg}
\end{table}
}

We also performed a more detailed investigation of 
$\log(\, g_{\One}(M,L)\, g_{\One}(M,L)\,)$, and 
$\log(\, g_{\One}(M,L)\, g_{\Phi(h)}(M,L)\,)$. These should both flow
from their UV values in table 1 to IR values of 0; but whereas there is a
single flow for the $(\One,\One)$ boundary conditions, 
$\log(\, g_{\One}(M,L)\, g_{\Phi(h)}(M,L)\,)$ depends on the
dimensionless parameter $h M^{-6/5}$.
For small $|h M^{-6/5}|$ this should still leave essentially a single
crossover region from $(\One,\Phi)$ conformal boundary conditions in
the UV to massive boundary conditions in the IR, but we should expect that 
for large $h M^{-6/5}$ there are two crossover regions -- first a
massless boundary flow at $L h^{5/6} \sim 1$ from the $(\One,\Phi)$
CBCs in the UV to effective $(\One,\One)$ CBCs, and then a crossover
from these CBCs to the  IR boundary conditions 
which should agree with the single
crossover for the true $(\One,\One)$ case.

In figure 2b we give plots of these functions
against $\log( M\! L )$  for various values of $(h M^{-6/5})$.
The two straight lines in this plot are the exact UV values from table 1.
The lowermost curve is $\log(\, g_{\One}(M,L)\, g_{\One}(M,L)\,)$
from BTCSA at 43 states, showing a smooth flow from the UV value $-0.64...$
to the IR value $0$.
The remaining curves are $\log(\, g_{\One}(M,L)\, g_{\Phi(h)}(M,L)\,)$
from BTCSA at 36 states. The values of $hM^{-6/5}$, 
starting with the uppermost line and ending with the lowest line,
are $-0.65$, $-0.5$, $-0.4$, $-0.25$, $-0.1$, 0, 0.1, 0.25,
0.5, 1.0, and 3.0 respectively.

{\begin{figure}[htb]
\[\begin{array}{c}
{\epsfxsize=.8\linewidth\epsfbox[0 50 288 238]{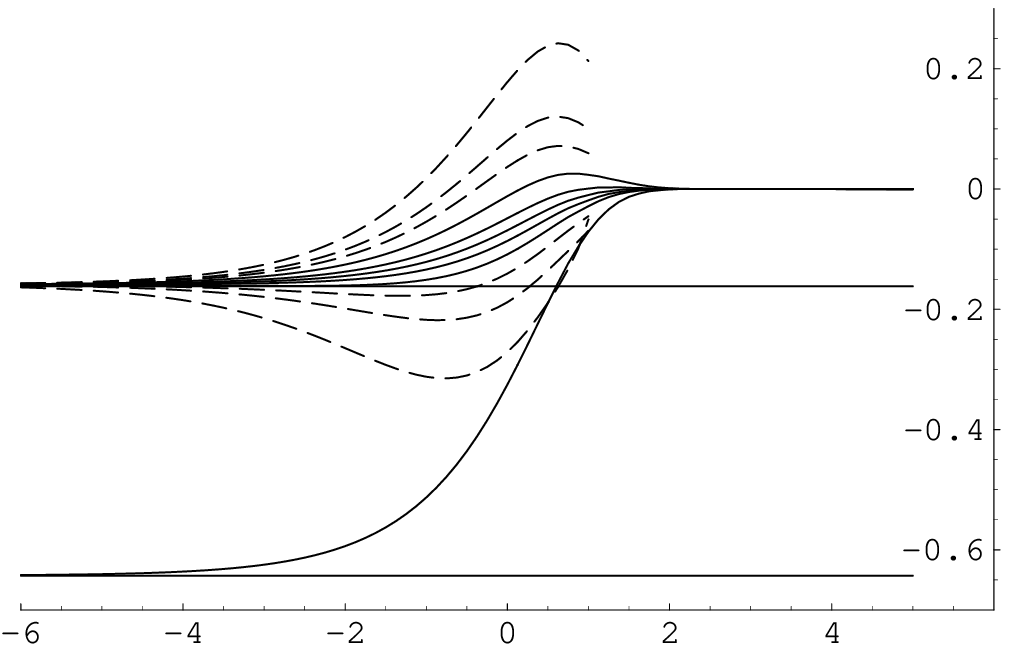}}\\
\parbox[t]{.8\linewidth}{\raggedright
\small 2b)
$\log(\, g_{\One}(M,L)\, g_{\One}(M,L)\,)$ and 
$\log(\, g_{\One}(M,L)\, g_{\Phi(h)}(M,L)\,)$ plotted against
against $\log( M\! L )$ for various values of $(h M^{-6/5})$;
details given in the text. Also shown are the exact UV values from
table 1.
}
\end{array}\]
\end{figure}}

There was one problem in finding numerical values for
$\log(\, g_{\One}(M,L)\, g_{\Phi(h)}(M,L)\,)$, and this was
estimating $B_{\alpha\beta}\, L$. It is only for large values of
$M\!L$ that equation (\ref{eq:aggb2}) holds, and for large values of 
$|h M^{-6/5}|$ this is outside the region of convergence of the BTCSA
method. We were only able to find $B_{\One\Phi(h)}\, L$ directly from 
equation (\ref{eq:aggb2}) for $|h M^{-6/5}| 
\lesssim 0.1$; for the remaining values of $h M^{-6/5}$, we
estimated $B_{\One\Phi(h)}\,L$  by extrapolation of the data for $|h
M^{-6/5}| \leq 0.0125$.

To mark this loss of accuracy, 
for $ -0.25\leq (h M^{-6/5}) \leq 0.25$
we have plotted $\log(\, g_{\One}(M,L)\, g_{\Phi(h)}(M,L)\,)$ for 
$-6< \log(M\!L) < 5$ with solid lines, but for the remaining values of 
$(h M^{-6/5})$ we have only given results for  $\log(M\!L) <1$, and
used dashed lines. We estimate the error in 
$\log(\, g_{\One}(M,L)\, g_{\Phi(h)}(M,L)\,)$ at $M\!L=1$ arising from
the extrapolation of $B_{\One\Phi(h)}\, L$ 
to be of the order of $0.01\%$ for $(hM^{-6/5})=0.3$, rising
to $20\%$ for $(hM^{-6/5})=3.0$.  
For larger values of $(hM^{-6/5})$ the errors in  extrapolation
render our data meaningless. 

For small values of $|h M^{-6/5}|$ (less than about $0.3$), we expect 
our results to be quite
accurate. While the quantitative results are less good for the larger
values,
we do still believe that there is good
qualitative agreement.
We now comment on the results in figure 2b.
There are essentially three different behaviours shown in this figure.

\begin{enumerate}

\item{}
For $ |h M^{-6/5}|\sim 0$ we found a single crossover to the massive
case, with only small deviations due to the boundary field.

\item{}
For $ (h M^{-6/5}) \gtrsim 1.0$, we found clear signs of two crossover
regions as $L$ varied from $0$ to $\infty$: the first (for small $L$)
a massless boundary flow from the value with 
$(\One,\Phi(0))$ b.c.s to that with $(\One,\One)$ b.c.s, and then a
second region (for larger $L$) with a crossover to zero as the bulk
mass scale dominates.
We were not able to go to large enough values of $ (h M^{-6/5}) $ to
truly separate the two crossover regions, but we think that figure
2b is very suggestive of the behaviour we propose.

\item{}
For some value of $(h M^{-6/5})$ between $-0.8$ and $-0.6$,
there is a critical value at which 
$\log(\, g_{\One}(M,L)\, g_{\Phi(h)}(M,L)\,)$ diverges -- this was
signalled by a pole in a rational fit to the data for
$B_{\One\Phi(h)}\,L$ for small $|h M^{-6/5}|$.
The predicted 
position of this pole is in agreement with the critical value 
$(h_{\rm crit}M^{-6/5}) = {-}0.68529$ discussed, from a different point of 
view, at the end of section~6 below.

\end{enumerate}

%
%
\resection{Boundary TBA}
For the rest of this paper we restrict our attention to
cases where the bulk is
massive, so that the bulk S-matrix is known.  In situations where 
boundary reflection factors are also known (or conjectured),
TBA-like equations for the ground-state energy on an interval
have been put forward
in refs.~\cite{Zta,LMSSa}. The derivation of these equations
begins with the expression given in~\cite{GZa} for the boundary state
$\ket{B_{\alpha}}$ corresponding to the boundary condition $\alpha$:
\eq
\ket{B_{\alpha}}=\exp\lf[\iintd
K_{\alpha}(\theta)A(-\theta)A(\theta)\ri]\ket{0}\,,
\label{bstate}
\en
where $K_{\alpha}(\theta)$ is related to the reflection factor for the
$\alpha$ boundary condition by $K_{\alpha}(\theta)=
R_{\alpha}(\frac{i\pi}{2}{-}\theta)$, and
$A(\theta)$ (denoted $A^{\dagger}(\theta)$ in \cite{LMSSa}) is the
Faddeev-Zamolodchikov operator creating a single particle with
rapidity $\theta$. (As in~\cite{LMSSa}, we ignore the possible presence
of additional
zero-momentum particles in this state. At least in some
regimes, this will be retrospectively justified via a comparison with
BTCSA data.)

The next step is to express the partition function
$Z_{\alpha\beta}(M,R,L)$ of the model on a cylinder of length $R$ and
circumference $L$ as
\eq
Z_{\alpha\beta}(M,R,L)\sim
{\phantom{|}}_L\!\bra{B_{\alpha}}\exp(-RH_{\rm p}(M,L))\ket{B_{\beta}}%
\!\!{\phantom{|}}_L
\en
where $H_{\rm p}(M,L)$ is the Hamiltonian for the system living on a circle of
circumference $L$, with periodic boundary conditions and bulk mass $M$, 
and $\ket{B_{\alpha}}\!\!{\phantom{|}}_L$ and
$\ket{B_{\beta}}\!\!{\phantom{|}}_L$ are boundary states set up as in
(\ref{bstate}), but on the circle rather than the
infinite line. Expanding the expressions for these states
and then making a
saddle-point approximation (bearing in mind the quantisation conditions
imposed on the momenta by the periodic boundary conditions) ultimately
leads to an expression for $-(\log Z_{\alpha\beta}(M,R,l/M))/l$ 
which becomes exact in the limit $l\rightarrow\infty$. But in 
this limit -- the opposite to that considered in the discussion following equation
(\ref{lrasympt}) --
the same quantity is given as $E_0^{\alpha\beta}(M,R)$,
the sought-after ground-state energy of $H_{\alpha\beta}(M,R)$.
The calculational details
can be found in ref.~\cite{LMSSa}; the upshot is that
$E_0^{\alpha\beta}(M,R)$ is expressed in terms of
the solution $\ep(\theta)$ to the following nonlinear integral
equation: 
\eq
\ep(\theta)=2r\cosh\theta-\log\lambda_{\alpha\beta}(\theta)-\phi{*}L(\theta)
\label{kermit}
\en
where $r=\MR$,
$L(\theta)=\log\bigl(1{+}e^{-\ep(\theta)}\bigr)$,
$f{*}g(\theta)=\frac{1}{2\pi}\iintd'f(\theta{-}\theta')g(\theta')\,$,
and
\eq
\lambda_{\alpha\beta}(\theta)=K_{\alpha}(\theta)K_{\beta}(-\theta)~,\qquad
\phi(\theta)=-i\prtial\log S(\theta)~,
\en
with $S(\theta)$ the bulk S-matrix~(\ref{asm}), and $M$ is the particle mass
(\ref{mlrel}).
The solution to~(\ref{kermit}) for
a given value of $r$ (and of any boundary-related parameters hidden inside
the labels $\alpha$ and $\beta$)
determines a function $c(r)$:
\eq
c(r)=\frac{6}{\pi^2}\iintd\, r\cosh\theta L(\theta)
\label{piggy}
\en
in terms of which $E_0^{\alpha\beta}(M,R)$ 
(which from here on we abbreviate to $E_0(R)$, when there is no danger
of confusion) is
\eq
E_0(R)=\Ebndry+\Eblk R-\frac{\pi}{24 R}c(\MR)~,
\label{bath}
\en
where $\Ebndry$ is a possible constant contribution to $E_0(R)$ coming
from the boundaries, and $\Eblk$ is the bulk energy per unit length.
We will also be working with a dimensionless ground state
scaling function $F(r)$. With the bulk mass now nonzero, the
discussion of BTBA results is most convenient if this is used to set
the overall length scale, rather than the boundary magnetic field used
in the definition (\ref{polenta}) of the last section. Thus we set
\eq
F(r)=\frac{R}{\pi}E_0(R)
=\frac{r}{\pi M}\Ebndry+\frac{r^2}{\pi M^2}\Eblk-\frac{1}{24}c(r)~.
\label{scooter}
\en

The equations (\ref{kermit})--(\ref{piggy})
are superficially very similar to the more familiar
TBA equations found for periodic boundary conditions, but there are 
some important new features, as we now show. 
The identity
$\lambda_{\alpha\beta}(\theta{-}i\pi/3)
\lambda_{\alpha\beta}(\theta{+}i\pi/3)=
\lambda_{\alpha\beta}(\theta)\,$ (which follows from the boundary
bootstrap equations for the scaling Lee-Yang model)
is enough to establish that the
function $Y(\theta)=\exp(\ep(\theta))$ solves the same $Y$-system as found
in the standard Lee-Yang TBA, namely~\cite{Zf} 
\eq
Y(\theta-i\pi/3)\,Y(\theta+i\pi/3)=1+Y(\theta)~.
\label{blob}
\en
{}From there, the standard periodicity
property $Y(\theta{+}5i\pi/3)=Y(\theta)$ follows by simple
substitution.  However the consequent
$r$-dependence of the solutions is different, because the
$\lambda_{\alpha\beta}(\theta)$ term in~(\ref{kermit}) gives 
the function
$\ep(\theta)$ a non-trivial behaviour near $\theta=0$, and this persists
even in the $r\rightarrow 0$ limit. The limiting shape of
$\ep(\theta)$ is 
a pair of `kink' regions near to $\theta=\pm\log(1/r)$,
separated by two plateaux from what might fancifully be called a
`breather' region near to $\theta=0$. (This is in contrast to the
situation in the usual TBA where there are just the two kink regions at
$\theta\approx\pm\log(1/r)$, separated by a single plateau of length
$2\log(1/r)$~\cite{Zb}.) As explained in \cite{Zf}, the $r$-dependence of 
the ultraviolet-limiting solutions creeps in via interactions between
the regions where the $\theta$-dependence of $\ep(\theta)$ remains
non-trivial, and here these are separated by a distance $\log(1/r)$, 
{\em half} the value found for periodic boundary conditions. As a result,
the regular part of the expansion of $c(r)$ is generically
a power series in
$r^{6/5}$, rather than the $r^{12/5}$ that is found when the boundary
conditions are periodic.

The other, `irregular', parts of the function $c(r)$ can be traced to the 
integral against $r\cosh\theta$ in (\ref{piggy}). The mechanism is as 
described in \cite{Zb}, though 
the arguments must be generalised to incorporate the effect of the
central breather region. If, in terms of the blocks $(x)$ defined in
equation (\ref{asm}), the reflection factors
$R_{\alpha/\beta}$ are equal to $\prod_{x\in A_{\alpha/\beta}}\!\usbl{x}\,$,
then the final result is that
\eq
c(r)=\frac{4\sqrt{3}}{\pi}\Biggl(\sum_{x\in A_{\alpha}\cup A_{\beta}}
\!\!\!\sin\frac{x\pi}{3}\Biggr)r-\frac{2\sqrt{3}}{\pi}r^2+
\hbox{`regular terms'},
\label{llm}
\en
where `regular terms' refers to the already-discussed power series in $r^{6/5}$.
As will be justified via a specific example in the next section, the
scaling function $F(r)$ is expected to expand in
powers of $r^{6/5}$ alone about $r=0$. This will reproduced by 
the BTBA result of (\ref{scooter}) and (\ref{llm}), so long as the irregular
terms in (\ref{llm}) cancel against the
explicit bulk and boundary energies included in (\ref{scooter}). This
requirement determines the constants $\Eblk$ and $\Ebndry$ 
exactly:
\eq
\Ebndry=\frac{1}{2\surd{3}}%
\Biggl(\sum_{x\in A_{\alpha}\cup A_{\beta}}%
\!\!\!\sin\frac{x\pi}{3}\Biggr)M
\qquad;\qquad
\Eblk=-\frac{1}{4\surd 3}M^2\,.
\label{beaker}
\en
The value of $\Eblk$ is the same as
is found when the boundary conditions are
periodic~\cite{Zb}. This is as it should be, since $\Eblk$
reflects a bulk property of the model.

Finally we remark that the driving term in~(\ref{kermit}),
$2r\cosh\theta-\log\lambda_{\alpha\beta}(\theta)\,$, is singular at
the poles and zeroes of $\lambda_{\alpha\beta}(\theta)$. Therefore a
set of $r$-independent points at which 
$Y(\theta)$ either vanishes or is infinite can be
read directly off~(\ref{kermit}), at least in the strip
$-\pi/3\le\im\theta\le\pi/3\,$\footnote{beyond this strip, the $Y$-system
relation (\ref{blob}) should be used.}. 
Points of the first sort, at which $L(\theta)$ is also singular,
are also seen in solutions of more usual TBA equations; they were
called `type II' in~\cite{DTb}. Those of the second sort are a new
feature of the boundary TBA, and will
be dubbed `type~III'.

\resection{The $(\One,\Phi(h))$ ground state in the BTBA}
We can now discuss the physical import of these results, with
particular reference to the strip with
$(\One,\Phi(h))$ boundary conditions mentioned in 
section~3. For this setup the Hamiltonian~(\ref{eq:tcsa1}) has
$\phi_{\rm L}=0$, $2{-}x_{\varphi}=12/5$, and $1{-}x_{\rm
R}=6/5$. Therefore, the conformal perturbation theory expansion of 
$F(r)$ does have a regular expansion in $r^{6/5}$, and so matches the
behaviour derived in the last section within the BTBA. 

Next, we should decide which reflection factors
describe the $(\One,\Phi(h))$ situation. 
We found that substituting
$\lambda_{\alpha\beta}(\theta)=K_{(1)}(-\theta)K_b(\theta)$
into the basic BTBA equation (\ref{kermit})
enabled us to reproduce BTCSA data for 
${-}0.68529<h<0.554411$,
the necessary values of $b$ lying in
the range $-2\le b\le 2$. The natural conclusion, which
will receive further support from its consistency with results to be
reported in the remainder of this paper, is to
associate boundary conditions with reflection factors as follows:
\bea
\One~~&\leftrightarrow&R_{(1)}(\theta)\label{waldorf}\\[2pt]
\Phi(h)&\leftrightarrow&\,R_b(\theta)\label{astoria}
\eea
The precise relation between $h$ and $b$ will be discussed shortly, but in
terms of $b$ the constant part of $E_0(R)$ at large $R$
now follows from the general result (\ref{beaker}),
and is:
\eq
\Ebndry=\frac{1}{2}\lf(\sqrt3-1+2\sin\frac{b\pi}{6}\ri)\!M~.
\label{gonzo}
\en

Our numerical work with this particular
$\lambda_{\alpha\beta}(\theta)$ was complicated by the presence,
for $-2<b<2$, of  a double type~II singularity in $L(\theta)$
at $\theta{=}0$. 
(At $b{=}{-}2$ there is no
singularity, and at $b{=}2$ it is of order $4$.)
Therefore, in the general case
$e^{-\ep(\theta)}$ is singular at $\theta{=}0$, and the direct
numerical integration of equation (\ref{kermit}) gives very unsatisfactory 
results: we estimated the accuracy for $c(r)$ at  small $r$ to be
between $10^{-2}$ and $10^{-3}$.
To alleviate this problem we defined
\eq
\hat{\ep}(\theta)= \ep(\theta)-  q \log \tanh 
\fract{3}{4}\theta+
\log\hat\lambda_{\alpha\beta}(\theta)~~,~~
\hat{L}(\theta)=  
\log\lf( \tanh^q\fract{3}{4} \theta
+ \hat\lambda_{\alpha\beta}(\theta) e^{-\hat{\ep}(\theta)}\ri)
\label{eq2}
\en
where $q$ is the order of the pole at the origin ($0$, $2$ or $4$),
\eq
\hat\lambda_{\alpha\beta}(\theta)=\lambda_{\alpha\beta}(\theta)
\lf(\frac{S(\theta{-}i\pi/3)}{S(\theta{+}i\pi/3)}\ri)^{q/2}\,,
\label{eq3}
\en
and $\alpha=(1)$, $\beta=b$ for this particular case.
(The motivation for these redefinitions came from
analogous man\oe uvres performed
in refs.~\cite{BLZa,DTb} for certain excited-state TBA equations
for periodic boundary conditions.)
Equations (\ref{kermit}) and (\ref{piggy}) can then be
recast in the following form:
\eq
\hat{\ep}(\theta)=2r\cosh\theta-\phi{*}\hat{L}(\theta)~~,~~
c(r)=
{12\sqrt3\over \pi}qr+
\frac{6}{\pi^2}\iintd\, r\cosh\theta\hat{L}(\theta)
\label{eq5}
\en
These revised equations are nonsingular on the real axis,
and so can be solved numerically with higher accuracy.

The attempt to go beyond the range $-2<b<2$ exposes a couple of
subtleties of the boundary TBA. We start with the situation as the
point $b=-2$ is approached.
A study of the singularities of $L(\theta)$ in the 
complex $\theta$-plane
revealed
a pair of $Y(\theta)=-1$ (`type I' in the language of~\cite{DTb})
singularities on
the imaginary axis, at 
$\theta=\pm\theta_p$ say. They are confined to the segment
$ 0 < |\im \theta_p| < (b{+}2)\pi/6 $
by the double type II singularity at the origin, and type III
singularities at $\pm i(b{+}2)\pi/6$.
As $b$ approaches $-2$ from above, the length of this segment shrinks
to zero and the points $\pm\theta_p$
are forced towards the origin. In order to continue round $-2$ to
smaller real values of $b$, a deformation of integration contours is
therefore required.
This is just as occurs during the analytic continuation 
(in $r$) of ordinary TBA equations, discussed in refs.~\cite{DTa,DTb}.
When the deformed contours are returned to the real
$\theta$-axis, residue terms are picked up~\cite{DTa}, and
the singularities at $\pm\theta_p$ become
`active', in that their positions appear explicitly in the 
analytically-continued
equations. These equations are:
\bea
&&\ep(\theta)=2r\cosh\theta
- \log\lambda_{(1)b}(\theta)
+ \log { S(\theta-\theta_p) \over S(\theta+\theta_p) }
-\phi{*}L(\theta)\,,
\label{sam}\\[2pt]
&&c(r)=\frac{6}{\pi^2}\iintd r\cosh\theta L(\theta)
+  i {24 r \over \pi} \sinh\theta_p\,. 
\label{eq7}
\eea
As in \cite{DTa}, we adopt the convention that $\theta_p$ has a
positive imaginary part {\it after}\ the real $\theta$ axis has been
crossed, and the corresponding
singularity has become active. Its precise value 
appears as a free parameter in the equations, and must be fixed
by imposing
\eq
\ep(\theta_p) =  i\pi\,.
\label{zoot}
\en
These equations describe the ground state, for all
real $r$, whenever $b$ is in the interval $-4<b<-2$. As before, the
redefinitions (\ref{eq2}) can be used to put the equations into a form
better suited to numerical work.

At $b{=}2$, a different phenomenon occurs: two (inactive)
type II singularities, at $\pm i(b{-}2)\pi/6$,
hit the real $\theta$ axis. As
explained in section 3.4 of ref.~\cite{DTb}, 
when in the standard TBA 
the real $\theta$ axis is crossed by 
type II singularities, the continued equations can always be recast
into the form that they had before the crossing, although in general
some of the active singularities may have to be relocated from their
analytically-continued positions.
The same is true here,
and the mechanism is as follows.
After the point $b{=}2$, the singularities at $\pm
i(b{-}2)$ are active and the initial
analytic continuation of the basic BTBA equation (\ref{kermit}) is
therefore
\eq
\ep(\theta)=2r\cosh\theta-\log\lambda_{(1)b}(\theta)
-\log\frac{S(\theta-i(b{-}2)\pi/6)}{S(\theta+i(b{-}2)\pi/6)}
-\phi{*}L(\theta)\,.
\en
Equation (\ref{sam}) arose in just the same way, though there was a
different sign for the extra term, because of the opposite signs of the
residues for type I and type II singularities. This time there
is no need for an equivalent of equation (\ref{zoot}), since the
relevant singularities owe their existence to the factor
$\lambda_{(1)b}(\theta)$ and so, as mentioned at the end of
section~4, their positions are fixed irrespective of any other details
of the solution. Now the identity 
\eq
K_b(\theta)
\frac{S(\theta-i(b{-}2)\pi/6)}{S(\theta+i(b{-}2)\pi/6)}
=K_{4-b}(\theta)
\en
can be used to rewrite the analytically-continued equation as
\eq
\ep(\theta)=2r\cosh\theta-\log\lambda_{(1)4-b}(\theta)
-\phi{*}L(\theta)\,.
\en
As promised, this has the same form as the basic BTBA equation
(\ref{kermit}), which applied
before the point $b{=}2$ was passed, the only change being that the
parameter $b$ has been replaced by $4{-}b$. 
It is straightforward to check that the expression (\ref{piggy}) 
for $c(r)$ behaves in an analogous fashion under the continuation,
and so the system of equations has a (somewhat hidden) symmetry about
$b{=}2$.
Furthermore, as described in a related context in \cite{DTb}, there is
no reason for the $b$-dependence to be in any way singular at this point:
it is best
thought of as a `coordinate singularity', with the apparently
discontinuous behaviour residing solely in the BTBA equations, and not
in the functions $Y$ and $c$ that they encode. If the
original equations had been rewritten using contours shifted away from the
real $\theta$
axis for all integrations, then the same functions would have
been recovered, but from a system for which the point $b{=}2$ had no
special status. 

Combining the symmetry about $b{=}2$ with the evident symmetry of the
equations about $b{=}{-3}$ leads to the 
conclusion that the ground-state BTBA is periodic in
the parameter $b$ with period $10$. This might be
a little surprising, since
from~(\ref{eq:rfb}) the periodicity in $b$ of the driving term
$\lambda_{(1)b}(\theta)$ itself is $12$, but is well confirmed by
our numerical results.
Consider now the `regular' part of $c$, a function of $b$ and $r$
with an expansion in powers of $r^{6/5}$. By the result
just established, each coefficient in this expansion
must be a periodic function of $b$ with period $10$.
An excellent numerical fit for the first coefficient turns out to be
$-5.1041654269(9) \sin((b{+}0.5)\pi/5)$ (remarkably, all higher modes
allowed by the periodicity appear to be absent). This 
coefficient can also be found in terms of $h$, and is equal to
$ 24\,{\Gamma(2/5)^{1/2}\,\Gamma(6/5)^{1/2}\,}{\Gamma(4/5)^{-1}\,}
          (\pi\,M)^{-6/5}\,h
 =1.6794178\,M^{-6/5}\,h\,$. 

Combining the two
expressions yields
the relation between $b$ and $h$:
\eq
h(b)={-}0.685289983(9)
     \sin\Bigl((b{+}0.5)\pi/5\Bigr)M^{6/5}\,.
\label{fozzie}
\en
With $h(b)$ given by this formula we found an excellent numerical
agreement between the BTBA and 
BTCSA data at all real values of $b$. A particularly striking check
came on setting $b{=}{-}0.5\,$: from (\ref{fozzie}),
this should correspond to the `pure'
$\Phi$ boundary condition, with no boundary field, and should 
therefore have an expansion in powers of $r^{12/5}$. Thus {\em all}\ odd
powers of $r^{6/5}$ (and not just the first)
should vanish in the regular expansion of $c(r)$
at this point, and within our numerical accuracy
that is indeed what we found:
\bea
c(r)|_{b=-0.5}&=&
0.4 + 
\fract{6(\sqrt3{-}1)(2{-}\sqrt2)}{\pi}\,r-\fract{2\sqrt3}{\pi}\,r^2\nn\\
&&~
{}-8.18{\times}10^{-12}\,r^{6/5} + 0.554031116\,r^{12/5} \nn\\
&&~  
{}+1.14{\times}10^{-9}\,r^{18/5} - 
0.0025228 r^{24/5}-
2.88{\times} 10^{-7}\,r^6  + \dots\qquad
\label{animal}
\eea 

\resection{The generalisation to excited states}
We now turn to the excited states. In principle, the analytic
continuation method of~\cite{DTa} can be used to derive the relevant
generalisations of the basic BTBA equations. We have yet to complete
this in detail, but knowledge of the method allowed us to make an
educated guess as to the form that the equations would take, which we
then verified by means of a direct comparison with BTCSA data.
For the
one-particle states on the strip, the equations turned out to
have the same form as those
found in~\cite{DTa} for {\em two}-particle states on a
circle. With $(\One,\Phi(h))$ boundary conditions they are:
\bea
&&\ep(\theta)=2r\cosh\theta-\log\lambda_{(1)b}(\theta)
+ \log {S(\theta-\theta_0) \over S(\theta-\bar{\theta}_0)}
+ \log {S(\theta+\theta_0) \over S(\theta+\bar{\theta}_0)}
-\phi{*}L(\theta)\,,\qquad\quad
\label{eq101}\\
&&c(r)=\frac{6}{\pi^2}\iintd r\cosh\theta L(\theta)
+i{24 r\over\pi}(\sinh\theta_0 - \sinh\bar{\theta}_0)\,,\\[2pt]
&&\ep(\theta_0)=(2n{+}1)\pi i~~,~~
\ep(\bar\theta_0)=-(2n{+}1)\pi i\,,\label{r1}
\eea
where, as before, $\lambda_{(1)b}(\theta)=K_{(1)}(\theta)K_b(-\theta)$.
As in the last section, we desingularised these equations before
studying them numerically. 
For real $r$, $\bar\theta_0$ is equal to $\theta_0^{\,*}$, 
the complex conjugate of $\theta_0$, and so only one of the conditions
(\ref{r1}) needs to be
imposed. This is the generic form of the equations for $-2<b<2$. 

For levels for which the one-particle excited-state
equations retain the form of (\ref{eq101})--(\ref{r1})
as
$r\rightarrow\infty$, a large-$r$ asymptotic can be extracted 
much as in~\cite{BLZa,DTa}. The result is
\eq
E(R)-E_0(R)= M \cosh \beta_0
\label{bbae}
\en
where $\beta_0=\re(\theta_0)$ satisfies
\eq
2r\sinh \beta_0 -  i
\log R_{(1)} (\beta_0)R_{b} (\beta_0)=2\pi k~,
\label{bba}
\en
and the value of the integer $k$ is fixed by a combination of
the quantisation
condition (\ref{r1}) and the sign of $(\im(\theta_0){-}\pi/6)\,$:
\eq
k=2n +\fract{1}{2}(1{-}{\rm sign}(\im(\theta_0){-}\pi/6))\,.
\en
Equation (\ref{bba}) is just the `boundary Bethe ansatz' (BBA)
quantisation condition for the rapidity
$\pm\beta_0$ of a single particle
bouncing between the two ends of the interval. 
Further analytic results can be obtained, but for the rest of this
section we will instead discuss some general features that emerge
as $b$ is varied. 

For $b$ in the range $-4<b<-2$, a man\oe uvre similar to that seen in 
the last section for the ground state in the same region
appears to be necessary, resulting in the
appearance of a further
pair of active singularities in the system, situated
on the imaginary axis. Even for real $r$, there are then two
independent singularity positions to be fixed, making the equations
harder to handle numerically. We therefore leave this regime to one
side for the time being, and move on to the range 
$-2<b<-1$. In this regime, the basic one-particle equations 
(\ref{eq101})--(\ref{r1}) hold for all real $r$, for all one-particle
levels.  Figures 3a and 3b
compare BTBA results (points) with the first few levels
found using the BTCSA (continuous lines), at $b=-1.5$,
$h{=}0.402803\,M^{6/5}$. Notice that the plotted points
do not cover all of the lines visible on the figure: those missed
correspond to states containing more than one particle in the infrared,
and are presumably
described by BTBA systems with more active singularities than the 
four (at $\pm\theta_0$ and $\pm\bar\theta_0$) present in 
(\ref{eq101})--(\ref{r1}).

{\begin{figure}[t]
\[\begin{array}{ll}
\epsfxsize=.45\linewidth
\epsfbox[0 50 288 238]{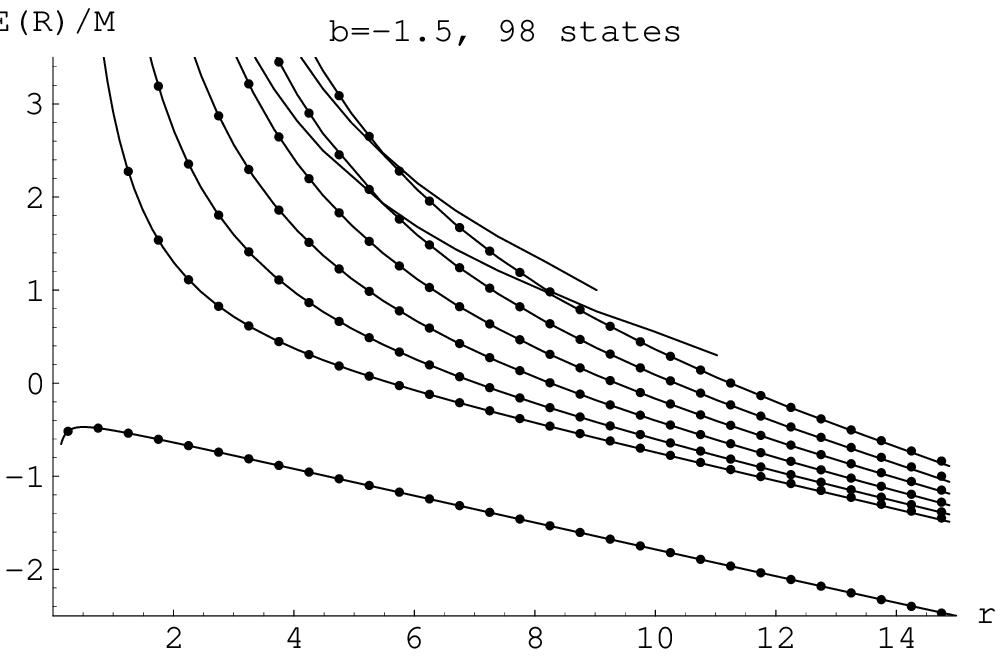}
{}~~&~~
\epsfxsize=.445\linewidth
\epsfbox[0 50 288 238]{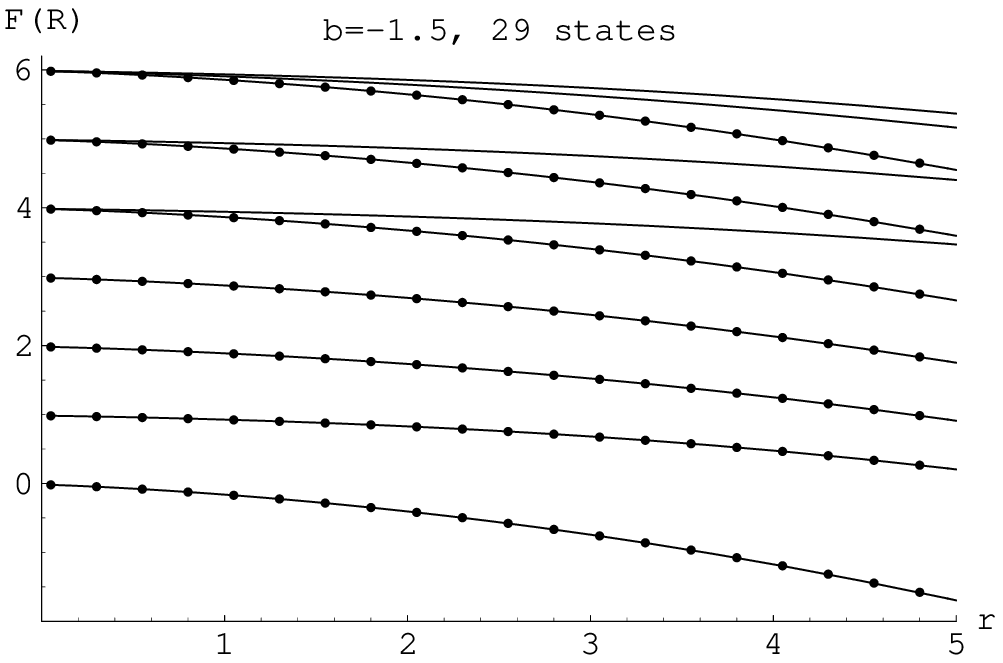}
\\
\parbox[t]{.45\linewidth}{\small 3a)
Energy levels $E(R)/M$ plotted against $r$ for the $(\One,\Phi(h))$
boundary conditions, with $b{=}{-}1.5$, 
$h{=}0.402803\, M^{6/5}$.
BTBA results (points) compared with the BTCSA (lines).
}
{}~~&~~
\parbox[t]{.45\linewidth}{\small 3b)
The same as figure 3a, but showing the scaling functions $F(r)$ to 
exhibit the ultraviolet behaviour.
}
\end{array}\]
\vskip -10pt
\end{figure}}

As $b$ passes $-1$, the lowest excited state as seen in plots of BTCSA
data breaks away from the other
excited levels and dips below the one-particle threshold.
Physically the reason for this is clear: at $b{=}{-}1$ an extra pole
in $R_b(\theta)$ enters the physical strip, signalling the appearance
of a boundary bound state. 
After this point the infrared behaviour of
this level changes from that of a free particle bouncing between 
the two boundaries to that of a particle trapped near to the $\Phi(h)$
boundary, a state with asymptotic gap $M\!\cos((b{+}1)\pi/6)$.
All of this can be seen in the BTCSA data, 
with the value of $b$ related to $h$ via~(\ref{fozzie}). Since
$h(b)$ had previously been obtained by a matching of data in
the ultraviolet, this infrared agreement
provides a nontrivial check on the consistency of our results.

{\begin{figure}[ht]
\[\begin{array}{ll}
\epsfxsize=.45\linewidth
\epsfbox[0 50 288 238]{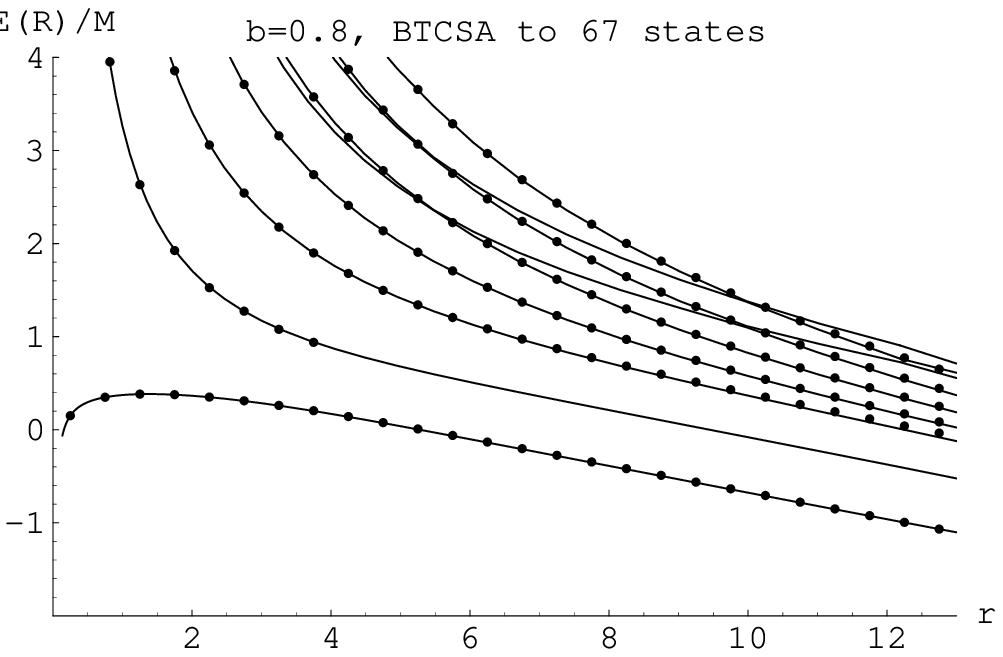}
{}~~&~~
\epsfxsize=.45\linewidth
\epsfbox[0 50 288 238]{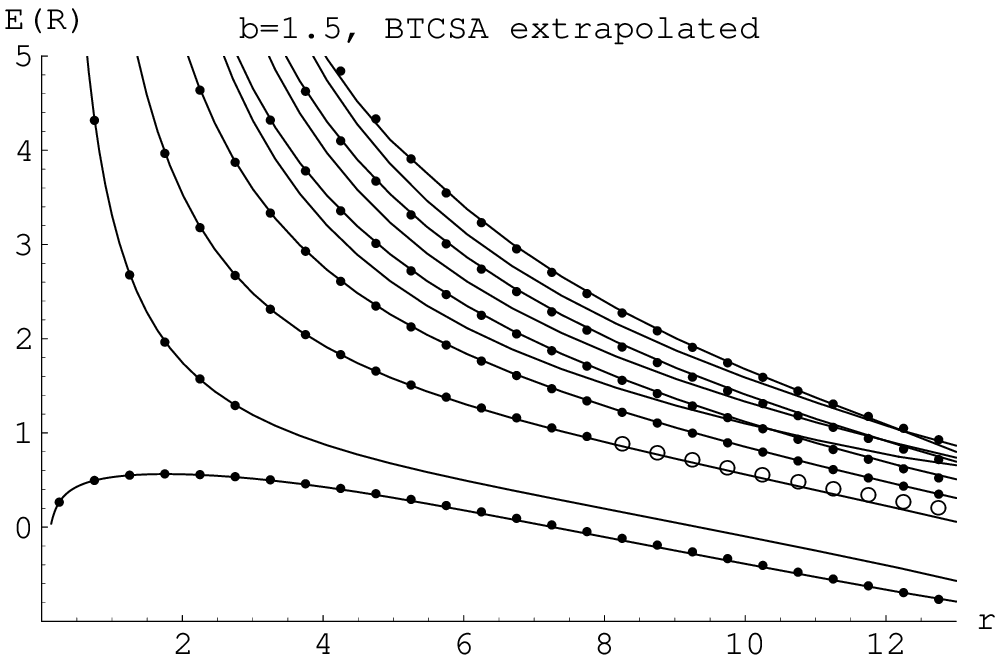}
\\
\parbox[t]{.45\linewidth}{\small 4)
Energy levels $E(R)/M$ plotted against $r$ for $b{=}{0.8}$, 
$h{=}{-}0.499555\,M^{6/5}$.  
Labelling as in figure 3a.
}
{}~~&~~
\parbox[t]{.45\linewidth}{\raggedright\small 5)
Energy levels $E(R)/M$ plotted against $r$ for $b{=}{1.5}$, 
$h{=}{-}0.65175 M^{6/5}$.   Labelling as in figure
3a, with BBA results plotted as ($\,\circ\,$).
}
\end{array}\]
\vskip -10pt
\end{figure}}

Figure 4 shows the situation at $b{=}0.8$,
$h{=}{-}0.499555\,M^{6/5}$, by which stage the dip in
the first level has become quite pronounced. The points on the graph
show BTBA results found using (\ref{kermit}) 
and (\ref{eq101})--(\ref{r1}), and it will be observed that the lowest set 
of excited points stops short. (The other sets persist to $r{=}\infty$
and for these the derivation of the BBA asymptotic (\ref{bbae})
is valid.) At $r\approx 4$, the basic 
excited-state BTBA (\ref{r1}) for the first excited level breaks
down, with an initial transition similar to that
observed in~\cite{DTb} for the second excited state of the $T_2$ model.
This reflects the fact that the particle is becoming bound to the
$\Phi(h)$ boundary. The BTBA equations become more complicated
in this regime, and we postpone their detailed study to future work. 
However in some regimes preliminary
predictions can be obtained, in the spirit of~\cite{KTWa}, by supposing
that solutions to the BBA equation~(\ref{bba}) can be continued
to complex values of $\beta_0$. This amounts to approximating the
problem by the quantum {\it mechanics}\ of a single particle
reflecting off the two walls with amplitudes $R_{(1)}$ and $R_b$.
For $-1<b<0$, the first excited level
as found from the BTCSA can be modelled with reasonable accuracy by a
solution $\beta_0(r)$ to the BBA which starts off real at 
small $r$, reaches zero at some ($b$-dependent) point $r_c$, and then
becomes purely imaginary, tending to $i (b{+}1)\pi/6$ from below
as $r$ tends to infinity. However for $b>0$ the only candidate solution is
satisfactory at best only for large $r$, which is why such points have
been omitted from figure~4. (The relevant $\beta_0(r)$
tends to $i(b{+}1)\pi/6$ from above at large $r$, implying an
approach to the asymptotic mass gap from below rather than above, but
our large-$r$ BTCSA data is not yet precise enough
to check on this prediction.)

Continuing to increase $b$, the next development occurs at $b{=}1$,
$h{=}{-}0.554411\,M^{6/5}$, 
where a second excited level starts to
drop below threshold, this time tending to a
gap of $M\!\cos((b{-}1)\pi/6)$ as $R\rightarrow\infty$. This heralds
the arrival of a second boundary bound state pole in $R_b(\theta)$
onto the physical strip.
Figure~5 
shows the situation for $b{=}1.5$. For this value of
$b$, the predicted dip in the second excited level is rather small
(about $0.03\,M\,$) and is only seen at 
rather large values of $R$, where
the BTCSA is at the limits of its useful range.
A more accurate
reflection of the infrared behaviour of this second excited level
is probably provided by the
analytically-continued BBA solution plotted on the figure as open
circles. The BTBA equations (\ref{r1}) now break down 
for both the first and the second excited levels, at $r\approx 4$ 
and $r\approx 9$ respectively.

Finally, at $b{=}2$, $h{=}h_{\rm
crit}{=}{-}0.68529\,M^{6/5}$,  the asymptotic mass gap
of the lowest excited state hits zero.  
It is not possible to increase $h$ any further
without making $b$ complex; if a larger value is used
in the BTCSA then we find energy levels becoming complex at large $r$. 
This can be seen analytically by combining (\ref{fozzie}) and 
(\ref{gonzo}) to write $\Ebndry$ as a function of $h\,$: a square-root
singularity is revealed at $h{=}h_{\rm crit}$, matching perfectly
behaviour that can be seen directly in the BTCSA.
Presumably, the boundary perturbation has destabilised the bulk
vacuum, a situation that will most probably repay further study.

%
%
\resection{Other combinations of boundary conditions}
Consideration of the strip with $(\One,\Phi(h))$ boundary conditions
has sufficed to establish the key relation (\ref{fozzie}), and has also
enabled us to check on the basic consistency of our associations
(\ref{waldorf}) and (\ref{astoria}) of the
reflection factors $R_{(1)}$ and $R_b$ with the $\One$ and
$\Phi(h(b))$ boundary conditions. However,
it is natural to wonder how the 
two other possible pairs of boundary conditions can be described, and
this is the topic of this section.

First, to the $(\One,\One)$ situation. From the identifications made
earlier, one might expect that ground state would be described by the
basic BTBA system (\ref{kermit}), with $\lambda_{\alpha\beta}(\theta)
=K_{(1)}(\theta)K_{(1)}(-\theta)$. However, a calculation of the
ultraviolet central charge as in \cite{LMSSa} yields the answer
$c(0)=2/5$, instead of the $-22/5$ expected on the basis of the
conformal result (\ref{rowlf}). Moreover, numerical comparisons of
BTBA and BTCSA results show no agreement. To motivate the resolution
of this problem, we  recall from section~3 that the $\One$ boundary
condition can be obtained from the $\Phi(h)$ boundary condition by
sending $h$ to ${+}\infty$. That was with the bulk massless, but if we
consider an interval of length $R$ with $(\One,\Phi(h))$ boundary
conditions and take the ${+}h/M^{6/5}\rightarrow\infty$ limit while
keeping $R$ of the same order as the bulk scale $1/M$, then it should
be possible (after suitable renormalisations) to arrive at the correct
BTBA.  We hope to say more about this elsewhere, but for now we just
observe, from (\ref{fozzie}), that at the level of the BTBA the
desired continuation will be found on setting $b=-3+i\hat b$, and then
varying $\hat b$ from zero to infinity.  The starting-point for this
continuation lies in the $-4<b<-2$ regime of the $(\One,\Phi(h))$
ground-state BTBA, and is hence described by  the modified system
(\ref{sam})--(\ref{zoot}).  This observation gave us the hint that the
$(\One,\One)$ ground state might also be described by a BTBA system in
which two singularities are already active. We therefore returned to
the numerical study of the modified equations
(\ref{sam})--(\ref{zoot}), this time with 
$\lambda_{\alpha\beta}(\theta) = K_{(1)}(\theta)K_{(1)}(-\theta)$.
The BTCSA data for the $(\One,\One)$ ground state was matched
perfectly over the full range of $r$, with a behaviour that is very
similar to that of the first excited-state TBA for the SLYM on a
circle~\cite{BLZa,DTa}. The equations exhibit a single transition
between an infrared regime, where two active singularities lie on the
imaginary $\theta$ axis, and an ultraviolet regime where these are
split and join the kink systems at $\theta\approx\pm\log(1/\MR)$. As a
result, each of these kink systems has a type~II singularity sitting
on the real $\theta$ axis, a fact which modifies the predicted value
of the ultraviolet effective central charge $c(0)$. The calculation is
just as described in refs.~\cite{BLZa,DTa}, and the desired result
$c(0)=-22/5$ is indeed recovered.  

For $(\Phi(h),\Phi(h'))$, calculations for the BTCSA are more
involved, since two irreducible Virasoro representations appear in
the decomposition (\ref{rowlf}) of $\cH_{(\Phi,\Phi)}$. Work on this is
still in progress, but in the meantime we have considered the one case
where a simple prediction can be made against which to test
conjectures for the BTBA, namely $h=h'=0$. With the boundary fields
set to zero, the regular part of $c(r)$ should exceptionally expand in
powers of $r^{12/5}$, rather than $r^{6/5}$, just as seen for the
$(\One,\Phi(0))$ boundary condition in (\ref{animal}).  For this case
we found unambiguously in favour of a BTBA of the modified type
(\ref{sam})--(\ref{zoot}), with
$\lambda_{\alpha\beta}(\theta)=K_{-0.5}(\theta)K_{-0.5}(-\theta)$. The
fit of $c(r)$ to the two irregular terms plus a series in $r^{6/5}$
for this proposal yields 
\bea
c(r)|_{b=b'=-0.5}&=&
0.4 - 
\fract{12(\sqrt3{-}1)(\sqrt2{-}1)}{\pi}\,r-\fract{2\sqrt3}{\pi}\,r^2\nn\\
&&~
{} - 3.89{\times}{{10}^{-12}}\,{r^{{\frac{6}{5}}}} +
1.79288\,{r^{{\frac{12}{5}}}}   \nn\\
&&~
{} - 4.48{\times}{{10}^{-9}}\,{r^{{\frac{18}{5}}}} 
 - 0.414179\,{r^{{\frac{24}{5}}}} - 9.54{\times}{{10}^{-7}}\,{r^6} 
 + \dots\qquad
\eea 
As in the earlier fit (\ref{animal}), the numerical data is consistent
with the coefficients of all odd powers of $r^{6/5}$ being zero, and
this supports the idea that the $(\Phi(0),\Phi(0))$ ground state is
indeed described by the modified BTBA system
(\ref{sam})--(\ref{zoot}). However this conclusion must remain
preliminary while a comparison with BTCSA data is lacking, and more
work will be needed before we can make any definitive statements about
the situation for other values of $h$ and $h'$.

%
%
\resection{Conclusions}
We hope to have shown that the combination of boundary truncated
conformal space and boundary thermodynamic Bethe ansatz techniques
allows a detailed analysis to be made of integrable boundary
models. Even for a theory as simple as the scaling Lee-Yang model, a
rich structure has been revealed. Our results support the contention
that the integrable boundary conditions of the model are $\One$, with
reflection factor $R_{(1)}(\theta)$, and the one-parameter family
$\Phi(h)$, with reflection factors $R_b(\theta)$. It is worth pointing
out that these latter are a particular subset of the reflection
factors given by Ghoshal in~\cite{Ga} for the lowest breather in the
sine-Gordon model, considered at the Lee-Yang point. It is natural to
suppose that a boundary variant of quantum group reduction is at work
here, and it would be interesting to explore this aspect
further. Perhaps even more natural would be to make a link with a
reduction of the boundary Bullough-Dodd model, since in that case the
classically integrable boundary conditions already lie in a couple of
one-parameter families~\cite{BCDRa}.

{}From BTCSA and BTBA results, and also from a simple consideration of
their ultraviolet limits, it is clear that the $\One$ boundary
condition is not the same as the $\Phi(h)$ boundary condition for any
finite value of $h$. Nevertheless, we note that
$R_{(1)}=R_{b=0}$. Thus the infrared data provided by an S-matrix and
reflection factor alone is not enough to characterise a boundary
condition completely. It is still possible that there is a special
relationship between the $\One$ and $\Phi(h(b{=}0))$ boundary
conditions, but further numerical work will be required before we can
make any definite claims one way or the other.

Finally we would like to reiterate that the boundary scaling Lee-Yang
model was studied in this paper just as a first example. The methods
that we have used should be applicable in many other situations, and
we intend to report on such matters in due course. 

%

\bigskip
\bigskip
\noindent{\bf Acknowledgements --- }
PED thanks Ed Corrigan,
and GMTW thanks M.~Blencowe, M.~Ortiz and I.~Runkel for
conversations, and A.~Honecker for pointing out reference
\cite{BWeh1}. We would also like to thank Jean-Bernard Zuber for
helpful comments. The work was supported in part by a TMR grant of the
European Commission, contract reference ERBFMRXCT960012, in part by a
NATO grant, number CRG950751, and in part by an EPSRC grant
GR/K30667. PED and GMTW thank the EPSRC for Advanced Fellowships, AJP
thanks the EPSRC for a Research Studentship, and RT thanks the
Mathematics Department of Durham University for a postdoctoral
fellowship and SPhT Saclay for hospitality.

\bigskip
\noindent{\bf Note --- } for some recent work exploring related
issues from a slightly different angle,
see ref.~\cite{LSSa}.
%
%
\bigskip
%
\renewcommand\baselinestretch{0.95}

%
%
\end{document}